\documentclass{article}

\input{psfig.sty}

\begin{document}

\title{CHIRAL UNITARY APPROACH TO HADRON SPECTROSCOPY\\
}

\author{E. Oset$^1$, T. Inoue$^1$, M. J. Vicente Vacas$^1$,
A. Ramos$^2$ and C. Bennhold$^3$\\
$^1$ Departamento de Fisica Teorica and IFIC, Universidad de Valencia,\\
Institutos de Investigacion de Paterna, 
 Valencia, Spain\\
$^2$ Departament d'Estructura i Constituents de la Materia,\\
 Universitat de
Barcelona, Diagonal 647,
Barcelona, Spain\\
$^3$Center for Nuclear Studies, Department of Physics,\\ 
The George
Washington University,
Washington D. C. 20052}

\maketitle

\begin{abstract}
The s-wave meson-baryon interaction in the $S = -1$, $S= 0$ and $S= -2$ sectors
is studied by means of coupled channels, using the lowest-order
chiral Lagrangian and the N/D method or equivalently the Bethe-Salpeter equation
to implement unitarity.  
 This chiral 
approach leads to the dynamical generation of the $\Lambda (1405)$,
$\Lambda(1670)$ and $\Sigma(1620)$ states for  $S = -1$, the $N^*(1535)$ for 
$S= 0$ and the $\Xi(1620)$ for $S= -2$. We 
look for poles in the complex plane and extract the
couplings of the resonances to the different final states. This allows identifying
 the $\Lambda (1405)$ and the $\Lambda(1670)$ resonances with $\bar{K}N$
and $K\Xi$ quasibound states, respectively. Our results are found to be
incompatible with the measured properties of the $\Xi(1690)$ resonance,
thus ruling this state out as the remaining
member of this octet of dynamically generated resonances.  
We therefore assign $1/2^-$ for the spin and parity of the $\Xi(1620)$ resonance as the 
$S=-2$ member of the lowest-lying $1/2^-$ octet.
                                              
\end{abstract}

The low-energy $K^-N$ scattering and transition to coupled channels is one of
the cases of successful application of chiral dynamics in the baryon
sector.
The studies of Refs.\cite{Kai95,Kai97} showed that one could
obtain an excellent description of the low-energy data starting from chiral
Lagrangians and using the multichannel Lippman-Schwinger equation to
account for multiple scattering and unitarity in coupled channels.  By including
all open channels above threshold and fitting a few chiral parameters of
the second-order Lagrangian one could obtain a good agreement with the data at 
low energies.  This line of work was continued in Ref.\cite{angels}, where all coupled
channels that could be arranged
from the octet of pseudoscalar Goldstone bosons and the entire baryon ground state 
octet were included .
 In Ref.\cite{angels} it was demonstrated that
using the Bethe-Salpeter equation (BSE) with coupled channels and using the 
lowest-order chiral Lagrangians, together with one cut off to regularize the
intermediate meson-baryon loops, a good description of all
low-energy data was obtained.  One of the novel features with respect to other approaches
using the BSE is that the lowest-order meson-baryon amplitudes, playing the role
of a potential, could be factorized on shell in the BSE, and thus the set of
coupled-channels integral equations became a simple set of algebraic equations,
technically simplifying the problem.  The justification of this procedure
was developed in the treatment of meson-meson interactions using
chiral Lagrangians and the N/D method\cite{nsd}.  One uses dispersion
relations and shows that neglecting the effects of the left-hand singularity
(also shown to be small there) one needs only the on-shell scattering matrix
from the lowest-order Lagrangian, and  the eventual effects of higher-order 
Lagrangians are accounted for in terms of subtractions in the dispersion
integrals. 
The N/D method has also been recently applied to study pion-nucleon
dynamics\cite{om00}.

    The work of Ref.\cite{angels} was reanalyzed recently\cite{joseulf} 
from the point of view of the N/D method and dispersion relations, leading
formally to the same algebraic equations found in Ref.\cite{angels}. There are also
technical novelties in the regularization of the loop function, which is done
using dimensional regularization in Ref.\cite{joseulf}, while it was regularized
with a cut off in Ref.\cite{angels}. 

   One of the common findings shared by all the theoretical approaches is
the dynamical generation of the  $\Lambda(1405)$ resonance which appears with
the right width, and at the correct position, with the choice of a cut off of
natural size. This natural generation from the interaction of the meson-baryon
system with the lowest-order Lagrangian allows us to identify that state as a
quasibound meson-baryon state. This would explain why ordinary quark models have
had so many problems explaining this resonance\cite{quarks}.

   In ordinary quark models the $\Lambda(1405)$ resonance would mostly be a
SU(3) singlet of $J^P=1/2^-$ and there would be an associated octet of
s-wave  excited 
$J^P=1/2^-$ baryons that would include the N*(1535), the  $\Lambda(1670)$,
the $\Sigma(1620)$ and a $\Xi^*$ state.  In the chiral approach one would also 
expect the appearance of such a nonet of resonances. In fact, it appears  
naturally in the approach of Ref.\cite{angels},
with a degenerate octet, when setting all the masses of the octet of stable  
baryons equal on one
side and the masses of the octet of pseudoscalar mesons equal on the other side. 
 Yet, to obtain this 
result it is essential that the coupled 
channels do not omit any of the channels that can be constructed from the
octet of pseudoscalar mesons and the octet of stable baryons.

The lowest-order Lagrangian involving the octet of pseudoscalar mesons and 
the $1/2^+$ baryons is given in Refs.\cite{Pi95,Eck95,Be95,Mei93}.

At lowest order in momentum, that we will keep in our study, the interaction
Lagrangian reads

\begin{equation}
L_1^{(B)} = < \bar{B} i \gamma^{\mu} \frac{1}{4 f^2}
[(\Phi \partial_{\mu} \Phi - \partial_{\mu} \Phi \Phi) B
- B (\Phi \partial_{\mu} \Phi - \partial_{\mu} \Phi \Phi)] > \ ,
\label{eq:lowest}
\end{equation}
where $\Phi$ and $B$ are the SU(3) matrices for the mesons and baryons,
respectively and the symbol $< >$ stands for the trace of the resulting SU(3)
matrix. The Lagrangian of Eq.~(\ref{eq:lowest})
leads to a common structure of the type
$\bar{u} \gamma^u (k_{\mu} + k'_{\mu}) u$ for the different channels, where
$u, \bar{u}$ are the Dirac spinors and $k, k'$ the momenta of the incoming
and outgoing mesons.

The lowest-order amplitudes for these channels are easily evaluated from 
Eq.~(\ref{eq:lowest}) and are given by
\begin{equation}
V_{i j} = - C_{i j} \frac{1}{4 f^2}(2\sqrt{s} - M_{Bi}-M_{Bj})
\left(\frac{M_{Bi}+E}{2M_{Bi}}\right)^{1/2} \left(\frac{M_{Bj}+E^{\prime}}{2M_{Bj}}
\right)^{1/2}\, ,
\label{eq:ampl2}
\end{equation}
with $E,E^{\prime}$ the energy of the initial, final baryon, 
and the matrix $C_{i j}$, which is symmetric, is given in Ref.\cite{angels}.

Note that the use of physical masses in Eq.~(\ref{eq:ampl2}) 
effectively introduces  some contributions of higher orders in the
chiral counting. In the standard chiral approach one would be using the average
mass of the octets in the chiral limit and higher order Lagrangians involving
SU(3) breaking terms would generate the mass differences. By introducing the
physical masses one guarantees that the phase space for the reactions,
thresholds and unitarity in coupled channels are respected from the beginning.

Ref.\cite{joseulf}, using the N/D method\cite{nsd} for this particular case,
proved that the scattering amplitude could be written by means of the
algebraic matrix equation 
\begin{equation}
T = [1 - V \, G]^{-1}\, V
\label{eq:bs1}
\end{equation}
with $V$ the matrix of Eq.~(\ref{eq:ampl2}) evaluated on shell, or
equivalently
\begin{equation}
T = V + V \, G \, T
\label{eq:bs2}
\end{equation}
with $G$ a diagonal matrix given by the loop function of a meson and a baryon
propagators.

One can see that Eq. (\ref{eq:bs2}) is just the Bethe-Salpeter equation 
but with the $V$ matrix factorized on shell, which allows one to extract the
scattering matrix $T$ trivially, as seen in Eq.~(\ref{eq:bs1}).

 The analytical expression for $G_l$ can be obtained from Ref.\cite{OOP97} using a
 cut off and from Ref.\cite{joseulf} using dimensional regularization. In this
 latter case one obtains 
\begin{eqnarray} 
G_{l} &=& i 2 M_l \int \frac{d^4 q}{(2 \pi)^4} \,
\frac{1}{(P-q)^2 - M_l^2 + i \epsilon} \, \frac{1}{q^2 - m^2_l + i
\epsilon}  \nonumber \\ &=& \frac{2 M_l}{16 \pi^2} \left\{ a_l(\mu) + \ln
\frac{M_l^2}{\mu^2} + \frac{m_l^2-M_l^2 + s}{2s} \ln \frac{m_l^2}{M_l^2} +
\right. \nonumber \\ & &  \phantom{\frac{2 M}{16 \pi^2}} +
\frac{\bar{q}_l}{\sqrt{s}} 
\left[ 
\ln(s-(M_l^2-m_l^2)+2\bar{q}_l\sqrt{s})+
\ln(s+(M_l^2-m_l^2)+2\bar{q}_l\sqrt{s}) \right. \nonumber  \\
& & \left. \phantom{\frac{2 M}{16 \pi^2} +
\frac{\bar{q}_l}{\sqrt{s}}} 
\left. \hspace*{-0.3cm}- \ln(-s+(M_l^2-m_l^2)+2\bar{q}_l\sqrt{s})-
\ln(-s-(M_l^2-m_l^2)+2\bar{q}_l\sqrt{s}) \right]
\right\} \ ,
\label{eq:gpropdr}
\end{eqnarray}        
which has been rewritten
in a convenient way to show how the imaginary part of $G_l$ is
 generated and how one can go to the unphysical Riemann sheets in order to identify
 the poles. 
The dimensional regularization scheme is preferable
if one goes to higher energies where the on-shell momentum of the intermediate
states is not much smaller than the cut off.

\section{Strangeness $S= -1$ sector}

We take the $K^- p$ state and all related channels using SU(3) mesons and baryons
within the chiral
approach, namely $\bar{K}^0 n$, $\pi^0 \Lambda$, $\pi^0 \Sigma^0$,
$\pi^+ \Sigma^-$, $\pi^- \Sigma^+$, $\eta \Lambda$, $\eta
\Sigma^0$, $K^0\Xi^0$ and
$K^+\Xi^-$.  Hence we have a problem with ten coupled channels.
The coupled set of Eqs.~(\ref{eq:bs1}) 
were solved in Ref.\cite{angels} using a cut off momentum of 630
MeV in all channels. Changes in the cut off can be accommodated in terms
 of changes in $\mu$, the regularization scale in the dimensional
 regularization formula for  $G_l$, or in the subtraction constant
$a_l$. In
 order to obtain the same results as in Ref.\cite{angels} at low energies, we set
 $\mu$ equal to the cut off momentum of 630 MeV (in all channels) and then
find the values of the
 subtraction constants $a_l$ such as to have $G_l$ with the same value
with the
 dimensional regularization formula (Eq.~(\ref{eq:gpropdr})) and the cut
off formula of \cite{angels}
at the $\bar{K} N$
 threshold. This determines the  values
\begin{equation} 
\begin{array}{lll} a_{{\bar K}N}=-1.84~~ &
a_{\pi\Sigma}=-2.00~~ & a_{\pi\Lambda}\,=-1.83 \\ a_{\eta
\Lambda}\,\,=-2.25~~ & a_{\eta\Sigma}=-2.38~~ & a_{K\Xi}=-2.52 \ . 
\end{array} 
\label{eq:coef}
\end{equation}
This guarantees that we obtain the same results at low energies as
in Ref.\cite{angels} and we find indeed that this is the case when
we repeat the calculation with the new $G_l$ of Eq.~(\ref{eq:gpropdr}).
Then we extend the results at
higher energies, looking for the possible appearance of new
resonances. 

\begin{figure}[ht]
\centerline{
\psfig{file=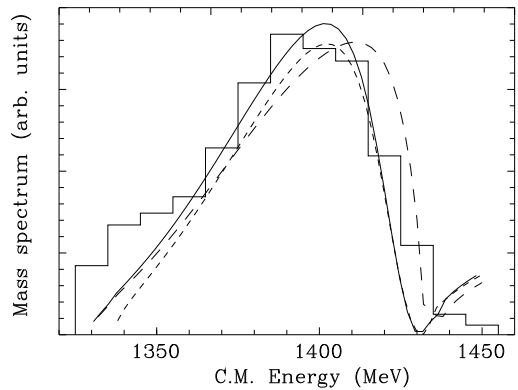,height=5.5cm,angle=0,silent=} 
}
\caption{The $\Lambda(1405)$ resonance obtained from the invariant
$\pi\Sigma$ mass distribution, with the full basis of physical
states
(solid line), omitting the $\eta$ channels (long-dashed line) and
with the isospin-basis (short-dashed line). 
\label{fig:lambda}}
\end{figure}

For the purpose of this study let us recall that Ref.\cite{angels} 
obtained the $\Lambda(1405)$ resonance which we show in Fig.~\ref{fig:lambda}
obtained from the $\pi \Sigma$ spectrum. Next we go to higher energies 
and search for new resonances.

\begin{figure}
\centerline{
\psfig{file=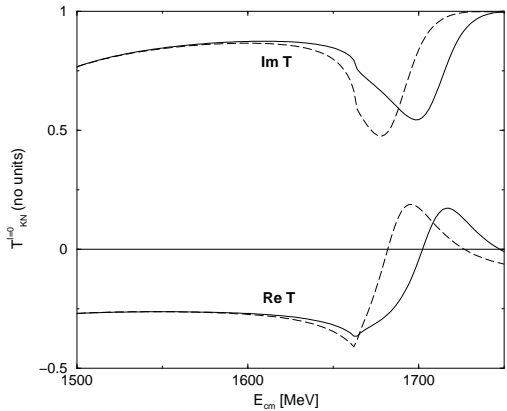,height=5.5cm,angle=270,silent=} 
}
\caption{Real and imaginary parts of the $\bar{K}N$ scattering
amplitude in the isospin $I=0$ channel in the region of the
$\Lambda(1670)$ resonance.
\label{fig:KN0}}
\end{figure}

Fig.~\ref{fig:KN0} shows the real and imaginary parts of the $I=0$
scattering amplitude, obtained in Ref.\cite{cornelius} normalized as in the
 Partial Wave Analysis
of Ref.~\cite{gopal77}. Remarkably, the amplitudes shown by 
the solid lines, which are obtained
using the low-energy parameters in Eq.~(\ref{eq:coef}),
show the resonant structure of the $\Lambda(1670)$
appearing at about the right energy and with a similar size
compared to the experimental analysis\cite{gopal77}.
The position of the resonance is quite
sensitive to the parameter $a_{K\Xi}$ and moderately sensitive to 
$a_{\eta \Lambda}$. Hence, without spoiling the nice agreement at low
energies, which is not sensitive to $a_{K\Xi}$, we 
exploit the freedom in the parameters by choosing $a_{K\Xi}=-2.70$, 
moving the resonance closer to its experimental
position (dashed lines).

SU(3) symmetry, partly broken here due to the use
of physical masses, demands a singlet and an octet of resonances.
Within $S=-1$, we have already identified
the singlet $\Lambda(1405)$ and the $I=0$ member of
the octet, the $\Lambda(1670)$. 
Since we found the partial decay widths and couplings 
of the $\Lambda(1405)$ to ${\bar K}N$ states
and the $\Lambda(1670)$ to $K\Xi$ states to be very large, one is naturally 
led to identify these two
resonances as a ``quasibound" ${\bar K}N$ and $K\Xi$ state,
respectively.

Searching for the $I=1$ member of the octet, 
we find that the $I=1$ amplitudes in our model are smooth and show no
trace of resonant behavior, in line with experimental observation.
To explore this issue further we conducted a search for the poles of
the ${\bar K}N \to {\bar K}N$ amplitudes in the second Riemann sheet
and find two poles in the $I=0$ amplitude
($1426 + {\rm i} 16, ~ 1708 + {\rm i} 21 $), corresponding to the
$\Lambda(1405)$ and the $\Lambda(1670)$, and one in the $I=1$ amplitude 
$(1579 + {\rm i} 296)$,
corresponding - most likely - to the resonance $\Sigma(1620)$. The large
width found for this resonance may explain why we saw no trace 
of it in the scattering amplitudes.

\section{Strangeness $S= 0$ sector}

  The strangeness $S= 0$ channel was also investigated using the Lippmann-
Schwinger equation and coupled channels in Ref.\cite{Kai95,siegel}. 
The $N^*(1535)$ resonance was found to be generated dynamically within this 
approach. Subsequently, work was done in this sector using the procedure 
of Ref.\cite{om00} with subtraction constants in Ref.\cite{Nacher:1999vg},
and the $N^*(1535)$ resonance, as well as the low-energy scattering 
observables, were well reproduced. The exception was the isospin 3/2 channel
which was not reproduced in this approach nor in 
\cite{Kai95,siegel}, and neither in \cite{Nieves:2001wt} where the $N^*(1535)$ 
resonance was generated together with the $N^*(1650)$ at the expense of 
using more free parameters.

 Ref.\cite{Inoue:2001ip} continued and further improved work along these lines
  by introducing the $\pi N \to \pi NN$ channels, which proved essential in
 reproducing the isospin 3/2 part of the $\pi N$ amplitude.

\begin{figure}[p]
\begin{center}
  \psfig{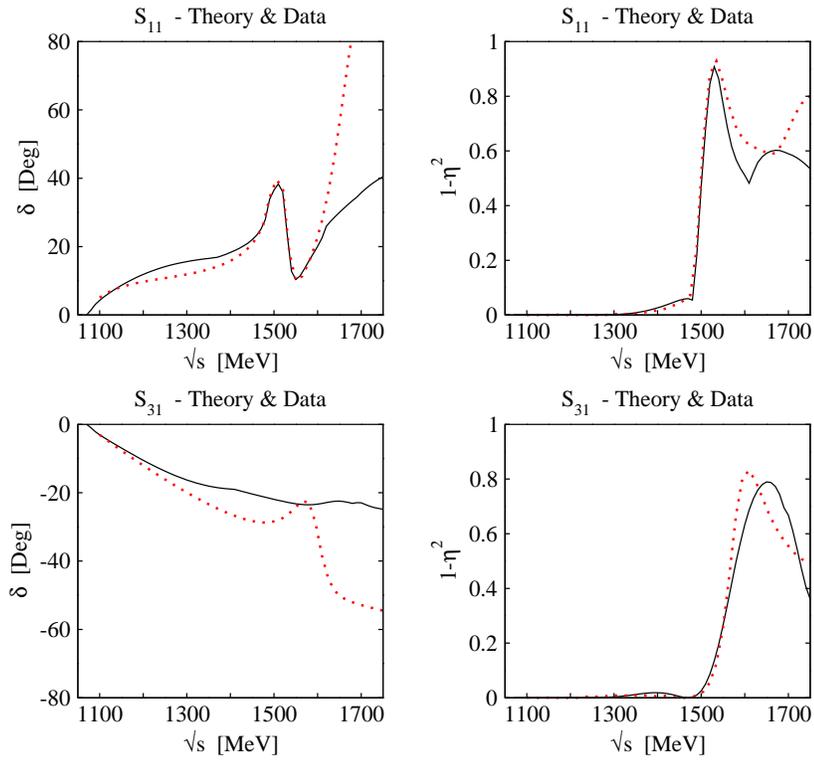}
  \caption{ 
    Phase-shifts and inelasticities of 
    $S_{11}$ and $S_{31}$ $\pi N$ scattering with $\pi \pi N$ channels.
          }
  \label{fig:phinfull}
\end{center}
\end{figure}

  For total zero charge one has six channels in this case,
 $\pi^- p$, $\pi^0 n$, $\eta n$, $K^+ \Sigma^-$,
 $K^0 \Sigma^0$, and $K^0 \Lambda$.  
  The subtraction parameters $a_i(\mu)$ for the meson-baryon propagators
 that we obtain in the new fit to the data, are
\begin{equation}
   \mu= 1200 ~\mbox{MeV},~~ 
   a_{\pi N}(\mu)     =  2.0 ,~~
   a_{\eta N}(\mu)    =  0.1 ,~~
   a_{K \Lambda}(\mu) =  1.5 , ~~ 
   a_{K \Sigma}(\mu)  = -2.8
\label{eqn:subpipin}
\end{equation} 
  The results obtained for the phase shifts and inelasticities of $\pi N$
  scattering are shown in Fig. \ref{fig:phinfull}, where the continuous line
  corresponds to the calculation while the dotted line is the experimental
  analysis.
  
 The $N^*(1535)$ peak is visible in the phase shifts and inelasticities of the
 $S_{11}$ amplitude in the panel.  The peak of the $\pi^- p \to \eta n$ cross
 section is also well reproduced but strength is missing just after the peak
 indicating the contribution of higher order partial waves. Figure 
 \ref{fig:phinfull} shows
 that the $S_{31}$ data are also fairly well reproduced once the $\pi N
 \to \pi \pi N $ channels are introduced in the approach. These amplitudes are
 fitted simultaneously to the scattering data and the  $\pi N \to \pi \pi N $
  cross sections and are somewhat different than those determined previously in 
 Ref.\cite{manley,burkhardt}.
 
\section{Strangeness $S= -2$ sector}

   Here we focus on the $S=-2$ sector for which,i.e., the
zero-charge states of the coupled-channels are 
$\pi^+ \Xi^-$, $\pi^0 \Xi^0$, $\bar{K^0} \Lambda$,
$K^-\Sigma^+$, $\bar{K^0} \Sigma^0$ and $\eta \Xi^0$. 

In the study of $S=-1$ resonances performed in Ref.\cite{cornelius}
the $a_l$ parameters were extracted by matching the results to
those of Ref.\cite{angels} and the range of values obtained, from
$-1.84$ to $-2.67$, serves as an indication for what we might assume
as reasonable natural size parameters
in the present $S=-2$ study.   
We search for poles in the second
Riemann sheet of the scattering amplitude, focussing on the elastic $\pi
\Xi \to \pi \Xi$ amplitude in the $I=1/2$ channel. 
As a trial run, we set the four values of the subtraction constants
to a value of $-2$ and
we discover a pole at $1607 + i 140$
MeV. This would lead to a width around 280 MeV, unacceptably large
compared to those of the two I=1/2 resonances of interest, the
$\Xi( 1620 )$ and the $\Xi(1690)$, which are reported to be of the
order of 50 MeV or less. The mass of the particle, around 1607
MeV, would be closer to the $\Xi(1620)$ resonance.

Allowing the subtraction constants $a_l$ to change within a reasonable
natural range, we obtain the results shown in Table \ref{tab:table2}. Only
$a_{\pi \Xi}$ and $a_{\bar{K}\Lambda}$ are
varied, since we find the couplings of the resonance to the $\bar{K}\Sigma$
and $\eta \Xi$ states to be very weak and therefore the results are
insensitive to the subtraction constants corresponding to these two
channels.
The values of the couplings, calculated from the residue of 
the diagonal scattering amplitudes \cite{cornelius},
are also shown in Table~\ref{tab:table2}. 

\begin{table}[ht]
\centering \caption{\small Resonance properties for various sets
of subtraction constants } \vspace{0.5cm}
\begin{tabular}{l|c|c|c|c|c|}
 & Set 1 & Set 2 & Set 3 & Set 4 & Set 5 \\
\hline
$a_{\pi\Xi}$          & $-2.0$ & $-2.2$ & $-2.0$ & $-2.5$ & $-3.1$ \\
$a_{{\bar K}\Lambda}$ & $-2.0$ & $-2.0$ & $-2.2$ & $-1.6$ & $-1.0$ \\
$a_{{\bar K}\Sigma}$  & $-2.0$ & $-2.0$ & $-2.0$ & $-2.0$ & $-2.0$ \\
$a_{\eta\Xi}$         &  $-2.0$ & $-2.0$ & $-2.0$ & $-2.0$ & $-2.0$ \\
\hline
$\mid g_{\pi\Xi}\mid^2$          & 8.7  & 7.2  & 7.4  & 7.2  & 5.9 \\
$\mid g_{{\bar K}\Lambda}\mid^2$ & 5.5  & 4.6  & 4.2  & 5.8  & 7.0 \\
$\mid g_{{\bar K}\Sigma}\mid^2$  & 0.68 & 0.59 & 0.54 & 0.74 & 0.93 \\
$\mid g_{\eta\Xi}\mid^2$         & 0.36 & 0.27 & 0.38 & 0.14 & 0.23 \\
 \hline
$M$ & 1607 & 1597 & 1596 & 1604 & 1605 \\  
$\Gamma/2$ & 140 & 117 & 134 & 98 & 66 
\end{tabular}
\label{tab:table2}
\end{table}

The second and third columns in Table~\ref{tab:table2} show that 
a change of 10\% in the subtraction constants
$a_{\pi \Xi}$ and $a_{\bar{K}\Lambda}$ modifies the mass of the
resonance only slightly but has a larger influence on the width.
Investigating the dependence of the results on the values of these
two subtraction constants we observe that the mass of the resonance is
confined to a range around 1600 MeV. The width, on the other hand,
can be reduced considerably
by a simultaneous increase of the strength of  $a_{\pi
\Xi}$ and a decrease of $a_{\bar{K}\Lambda}$, while
keeping both of them negative and still reasonably close
to the reference value of $-2$. In the last 
column we see that the width can be reduced to 130
MeV with acceptable values for the coefficients. While this width
might still appear as grossly overestimating the experimental ones,
we show below that this is not the case.

\begin{figure}[tbp]
\begin{center}
\psfig{width=6cm,angle=-90,file=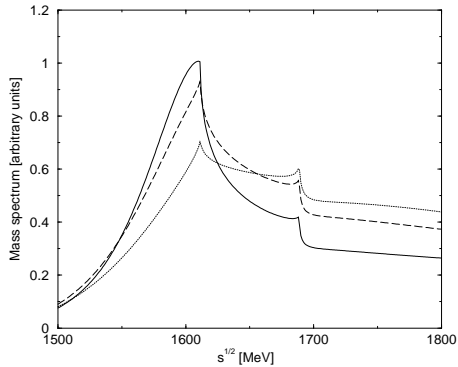}
\caption{The $\pi\Xi$ invariant mass distribution as a function of the
center-of-mass energy, for several sets of subtraction constants.
Solid line: $a_{\pi \Xi}=-3.1$ and $a_{\bar{K}\Lambda}=-1.0$;
Dashed line: $a_{\pi \Xi}=-2.5$ and $a_{\bar{K}\Lambda}=-1.6$;
Dotted line: $a_{\pi \Xi}=-2.0$ and $a_{\bar{K}\Lambda}=-2.0$. The
value of the remaining two other subtraction constants, $a_{{\bar K}\Sigma}$
and $a_{\eta\Xi}$, is fixed to $-2.0$ in all curves.}
\label{fig:masspixi}
\end{center}
\end{figure}

Since the 
$\Xi(1620)$ resonance decays only into 
 $\pi \Xi$ final states, it is experimentally visible through
the $\pi \Xi$ invariant mass distribution 
in reactions leading, among others, to $\pi$ and $\Xi$
particles. Our calculated distribution, displayed
in Fig.~\ref{fig:masspixi}, 
shows a very interesting feature, namely a smaller
apparent width  compared to the one obtained
at the pole position. 
For instance, for the values of the
subtraction constants in the last column of Table~\ref{tab:table2} we
see in Fig.~\ref{fig:masspixi} (solid line) an
apparent Breit-Wigner width of around 50 MeV and a shape for the
distribution which resembles the experimental peaks observed.
This well-known phenomenon, usually referred to as Flatt\'e effect
\cite{flatte}, is due to the presence of a
resonance just below the threshold of a channel to which the
resonance couples very strongly. In our case the
 $\bar{K} \Lambda$ channel opens at 1611 MeV and,
as shown in Table~\ref{tab:table2}, the resonance couples
very strongly to that state.
What actually happens is that at an invariant energy close
to the resonance mass the amplitude is given essentially by the
inverse of the resonance width. As soon as the
threshold is crossed, the new channel leads to 
an additional energy-dependent contribution for the width which
grows very rapidly with increasing energy. This produces a fast
fall-off for the amplitude, leading to an apparent width much smaller
than the actual width at the pole. This phenomenon has been observed,
e.g., in the case of the $a_0(980)$ meson
resonance as discussed in Refs.~\cite{OOP97,nsd}.

The question now arises which of the two $I=1/2$ candidates
should be identified with the resonance obtained here. 
The value found for the mass of the state would suggest identification
with the $\Xi(1620)$ which is rated as a one-star resonance in the PDG and has
unknown spin and parity.
The $\Xi(1690)$ state is better known and is rated as a 3-star
resonance.  Even if the spin and parity are unknown, there is far
more information available for this resonance than for the $\Xi(1620)$
\cite{pdg}. 
Ref. \cite{dionisi} gives ratios of partial decay widths having
sufficient accuracy for us to draw conclusions from
the properties of the $\Xi$ resonance found in this work.
We therefore investigate whether the parameters of the theory 
provide enough flexibility
to produce a pole with a real part closer to 1690 MeV,              
since the results of Table~\ref{tab:table2} show that, by decreasing the
size of $a_{\pi \Xi}$
or $a_{\bar{K}\Lambda}$, one increases the mass of the resonance.
However, the presence of the $\bar{K} \Lambda$ threshold leads to
mass values that stabilize around the cusp of this threshold
for a certain range of the parameters.
Continuing to change the $a_l$ parameters beyond this range 
does not increase the resonance mass but 
leads to a disappearance of the pole -- and with it the resonance.
The above argument clearly favors identifying 
the resonance found here with the $\Xi(1620)$ state.

The other argument in favor of the $\Xi(1620)$ assignment is the
following:  The results of Table~\ref{tab:table2} show that the
resonance couples strongly to the $\pi
\Xi$ and the $\bar{K}\Lambda$ channels but very weakly to $\bar{K}\Sigma$
and $\eta \Xi$. This is opposite to the observed properties of
the $\Xi(1690)$ resonance, for which Ref.~\cite{dionisi} gives a ratio
of branching ratios for $\bar{K}\Sigma$ to $\bar{K}\Lambda$ around
3 and for $\pi \Xi$ to $\bar{K}\Sigma$ of less than 0.09. In our  
opinion, this
argument rules out identifying the resonance found here
with the $\Xi(1690)$ state.

    In summary,  we have demonstrated that the chiral approach 
 to the $\bar{K}N$ and the other coupled channels, which proved so successful at
 low energies, extrapolates smoothly to higher energies and provides the basic
 features of the  scattering amplitudes, generating the resonances which would
 complete the states of the  nonet of the $J^P=1/2^-$ excited states.  The
qualitative
 description of the data without adjusting any parameters is telling us that the
 basic information on the dynamics of these processes is contained in the chiral
 Lagrangians. There is still some freedom left with the chiral symmetry 
 breaking terms. In 
our formulation they would go into the $a_l$ subtraction constants, and the
use
of different decay constants for each meson, by means of which one could 
obtain a
better description of the data. 
The analysis of the poles and the couplings of
 the  resonances to the different channels lead us to identify
 the strong coupling of
 the $\Lambda(1405)$ resonance to the $\bar{K}N$ state and the large coupling 
 of the $\Lambda(1670)$ resonance to the $K\Xi$ state, allowing us to classify
 these resonances as quasibound states of $\bar{K}N$ and $K\Xi$, respectively.
 In the $S= -2$ sector, the study performed here has allowed us to identify the
 resonance generated dynamically with the $\Xi(1620)$, hence providing a prediction for
  the spin and parity of this resonance which is currently not assigned by PDG.
 
\subsection*{Acknowledgments}

 This work is 
partly supported by DGICYT contract numbers BFM2000-1326, PB98-1247,
by the EU TMR network Eurodaphne, contract no. ERBFMRX-CT98-0169, and
by the US-DOE grant DE-FG02-95ER-40907.

\end{document}